\documentclass[a4paper,11pt]{article}
\usepackage{pos}

\title{Prompt and non-prompt charm baryons with ALICE}

\manuallySeparateAuthors
\author*[a,b]{Jianhui Zhu}
\author{ for the ALICE Collaboration}

\affiliation[a]{Institute of Particle Physics, Central China Normal University,\\
  No.152 Luoyu Road, Wuhan, China}

\affiliation[b]{Istituto Nazionale di Fisica Nucleare, Sezione di Padova,\\
  Via Marzolo 8, 35131 Padova, Italy}

\emailAdd{jianhui.zhu@cern.ch}

\abstract{
Recent measurements of prompt charm-baryon production at midrapidity in pp and p--Pb collisions show baryon-to-meson yield ratios significantly higher than those in $\rm e^+e^-$ collisions, suggesting that the charm fragmentation is not universal across different collision systems. Thus, more precise and broader measurements of prompt charm-baryon production are crucial to study the charm quark hadronization in a partonic-rich environment like the one produced in pp collisions at the LHC energies. Prompt charm baryon-to-meson yield ratios in proton--nucleus collisions provide important information about possible additional modification of hadronization mechanisms, on cold nuclear matter effects, and on the possible presence of collective effects that could modify the production of heavy-flavor hadrons. The non-prompt charm-hadron production can provide information about the beauty sector and can be used to study flavor dependence of heavy-quark hadronization. In this contribution, the most recent results on prompt and non-prompt charm-hadron yield ratios in pp and p--Pb collisions and on charm fragmentation fractions and charm production cross section at midrapidity per unit of rapidity measured by ALICE are discussed.}

\FullConference{%
  41st International Conference on High Energy physics - ICHEP2022\\
  6-13 July, 2022\\
  Bologna, Italy
}

\begin{document}
\maketitle

\section{Introduction}
The production of heavy-flavor hadrons in high-energy hadronic collisions can provide important tests of the theory of quantum chromodynamics (QCD). The production cross sections of heavy-flavor hadrons can be calculated using the factorization approach as a convolution of three factors \cite{COLLINS198637}: the parton distribution functions (PDFs) of the incoming protons, the hard-scattering cross section at partonic level, calculated as a perturbative series in powers of the strong coupling constant $\alpha_{\rm s}$, and the fragmentation functions of heavy quarks to given heavy-flavor hadron species. The latter term parametrizes the result of the hadronization process, which is an inherently non-perturbative process related to, or even driven by, the confining property of QCD. The heavy-flavor baryon-to-meson yield ratio is an ideal observable to probe the hadronization mechanism since the contributions from parton distribution function and hard-scattering cross section terms almost cancel in the ratio. The $\rm \Lambda_c^+/D^0$ ratio in pp and p--Pb collisions at the LHC is enhanced with respect to measurements at the electron colliders, and predictions of models and event generators tuned on $\rm e^+e^-$ and $\rm ep$ experiments, suggesting that the charm fragmentation functions are not universal among different collision systems \cite{ALICE:2017thy, ALICE:2020wfu, ALICE:2021rzj}. Several hadronization mechanisms, such as colour reconnection (CR) beyond the leading colour approximation \cite{Christiansen:2015yqa}, quark coalescence \cite{Minissale:2020bif, Song:2018tpv}, and statistical hadronization model (SHM) \cite{He:2019tik} including a set of higher-mass charm-baryon states as prescribed by the relativistic quark model (RQM), have been proposed to explain this enhancement. The newest measurements of the prompt charm baryons $\rm \Lambda_c^+$, $\rm \Sigma_c^{0, ++}$, $\rm \Xi_c^{0, +}$, $\rm \Omega_c^{0}$, and non-prompt $\rm \Lambda_c^+$ performed with the ALICE experiment are presented in this contribution and used to verify predictions from these hadronization mechanisms.

\section{Prompt and non-prompt $\rm D^+/D^0$ and $\rm D_s^+/(D^0+D^+)$ yield ratio in pp collisions}
The ratios of prompt and non-prompt D-meson yield ratios $\rm D^+/D^0$ and $\rm D_s^+/(D^0+D^+)$ are measured at midrapidity as a function of $p_{\rm T}$ in pp collisions at $\sqrt{s}=5.02$~TeV with the ALICE experiment \cite{ALICE:2021mgk}. The results are independent of $p_{\rm T}$ within the current experimental precision and are compatible with the FONLL \cite{Cacciari:2012ny} predictions in the case of prompt $\rm D^0$ and $\rm D^+$ mesons and FONLL+PYTHIA 8, to model the B hadron decay kinematics, in the case of non-prompt D mesons. These predictions are based on the factorization approach with universal fragmentation functions.

\section{Prompt and non-prompt $\rm \Lambda_c^+/D^0$ yield ratio in pp collisions}

\begin{figure}
\centering
\includegraphics[width=0.393\textwidth]{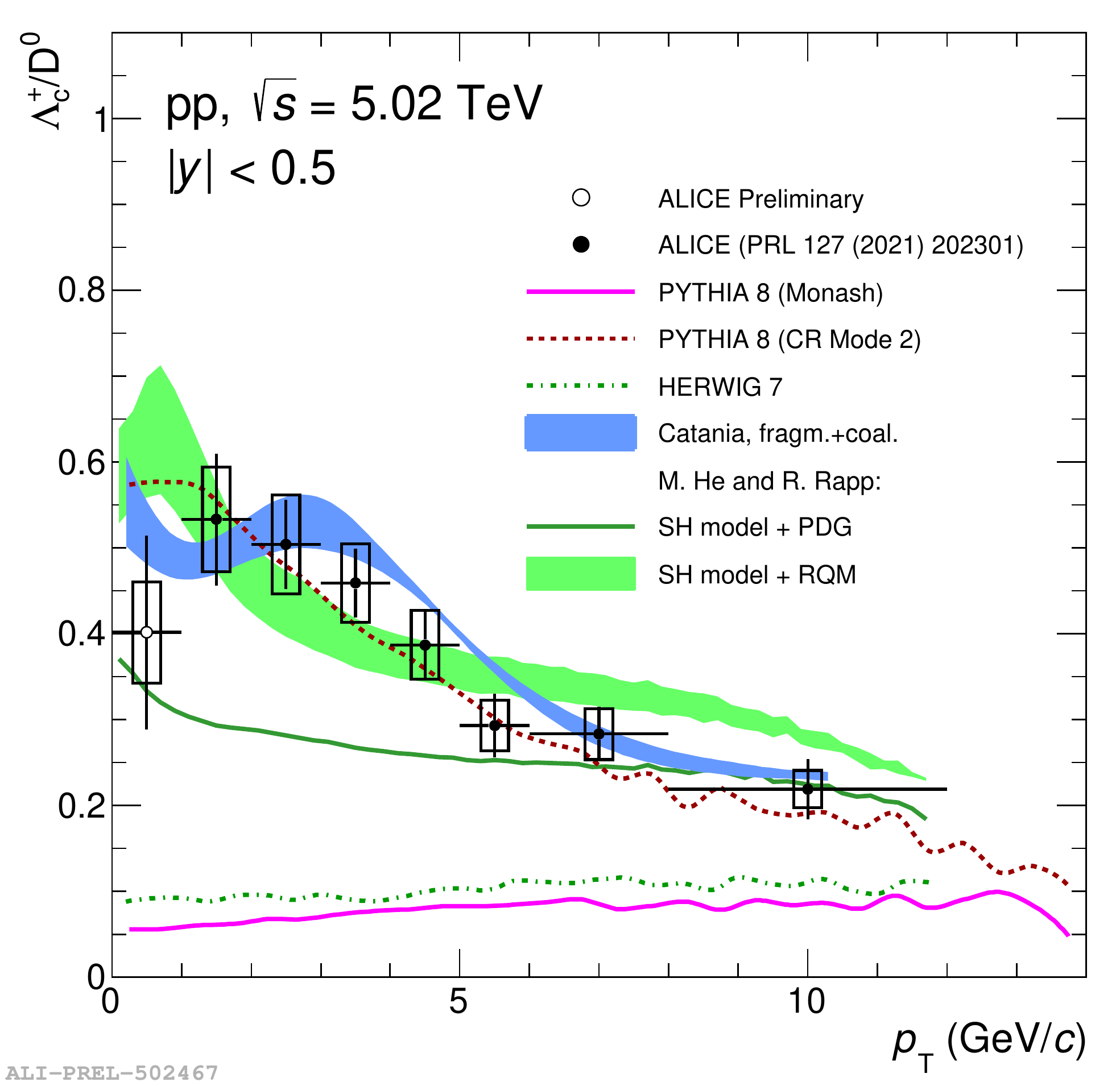}
\includegraphics[width=0.4\textwidth]{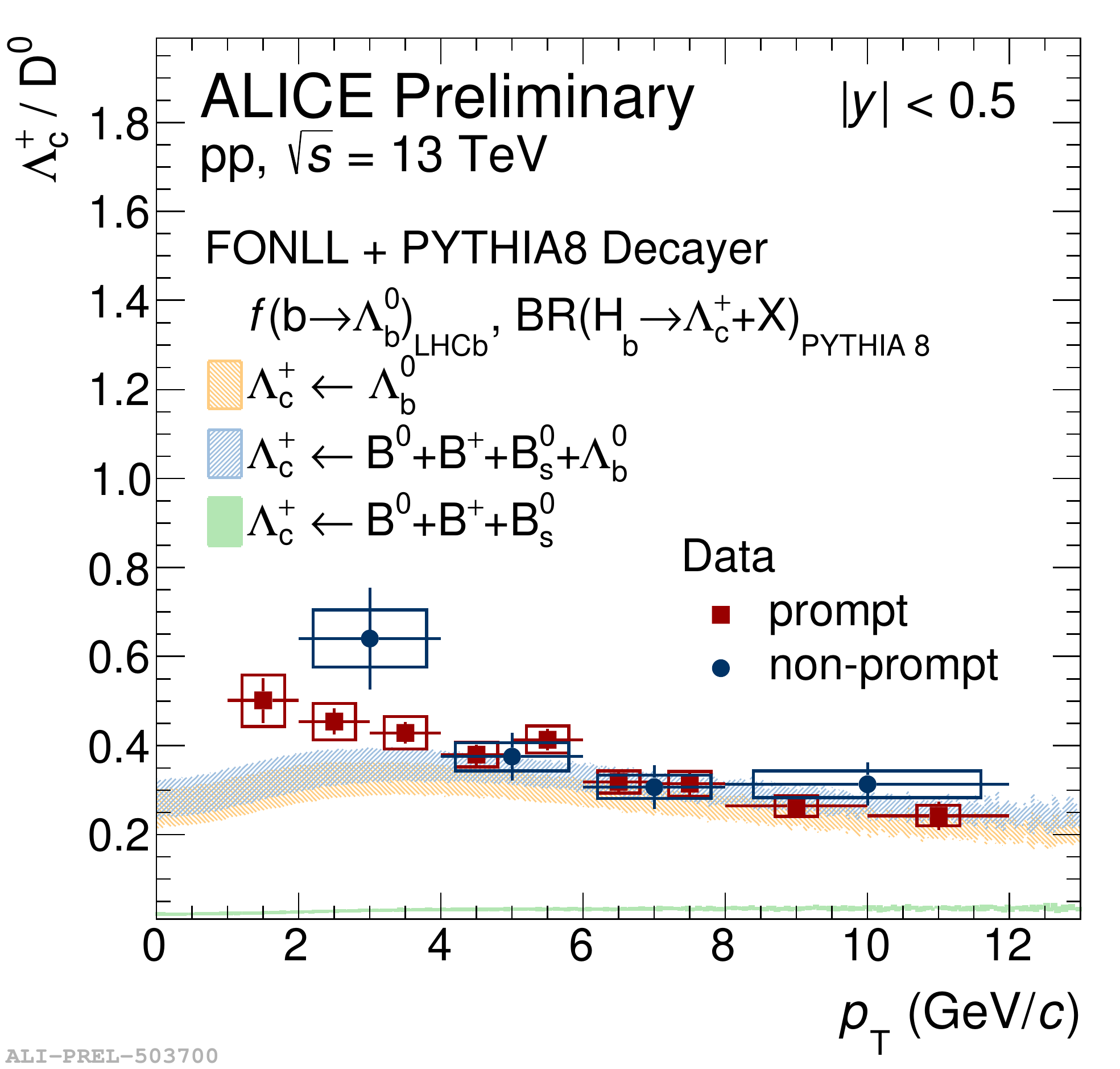}
\caption{Left: Prompt $\rm \Lambda_c^+/D^0$ yield ratio in pp collisions at $\sqrt{s}=5.02$~TeV compared with model predictions. Right: Prompt and non-prompt $\rm \Lambda_c^+/D^0$ yield ratio in pp collisions at $\sqrt{s}=13$~TeV compared with model predictions.}
\label{fig_1}       
\end{figure}

In Fig.~\ref{fig_1}, the prompt $\rm \Lambda_c^+/D^0$ yield ratio measured at midrapidity as a function of $p_{\rm T}$ in pp collisions at $\sqrt{s}=5.02$~TeV (left panel) and at $\sqrt{s}=13$~TeV (right panel) is shown. Recently, the production cross section of prompt $\rm \Lambda_c^+$ has been extended down to $p_{\rm T}=0$ at both collision energies. The prompt $\rm \Lambda_c^+/D^0$ is significantly underestimated by the Monte Carlo generator PYTHIA 8 (Monash tune) \cite{Skands:2014pea} tuned on measurements in $\rm e^+e^-$ collisions, but it is better described by a model with color reconnection (CR) beyond the leading color approximation \cite{Christiansen:2015yqa}, a statistical hadronization model with an augmented set of charm baryon states predicted by the relativistic quark model (SHM+RQM) \cite{He:2019tik}, or a model relying on hadronization via coalescence and fragmentation (Catania) \cite{Minissale:2020bif}.

The non-prompt $\rm \Lambda_c^+/D^0$ yield ratio measured in pp collisions at $\sqrt{s}=13$~TeV is shown in the right panel of Fig.~\ref{fig_1}. It is similar to the prompt $\rm \Lambda_c^+/D^0$ yield ratio. The data points are compared with theoretical models based on the B-hadron cross section predicted by FONLL and the beauty-quark fragmentation fractions to $\rm \Lambda_b^0$ baryons measured by the LHCb collaboration at forward rapidity \cite{LHCb:2019fns}, $f(\rm b\rightarrow\Lambda_b^0)$, and to B mesons in $\rm e^+e^-$ collisions \cite{10.1093/ptep/ptac097}, $f(\rm b\rightarrow B)$. The resulting beauty-hadron cross section was then folded with the $\rm H_b \rightarrow H_c + X$ decay kinematics obtained with PYTHIA 8 in order to obtain the non-prompt $\rm \Lambda_c^+$ cross section. The data and the model are compatible within 1.5$\sigma$.

\section{Prompt $\rm \Sigma_c^{0, +, ++}$, $\rm \Xi_c^{0, +}$ and inclusive $\rm \Omega_c^0$ over $\rm D^0$ yield ratios in pp collisions}

\begin{figure}
\centering
\includegraphics[width=0.307\textwidth]{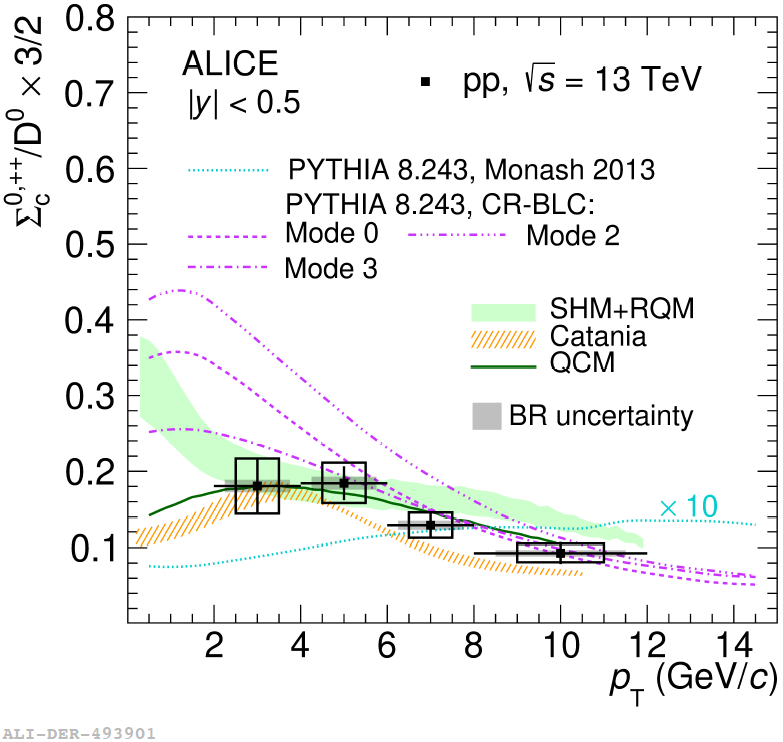}
\includegraphics[width=0.33\textwidth]{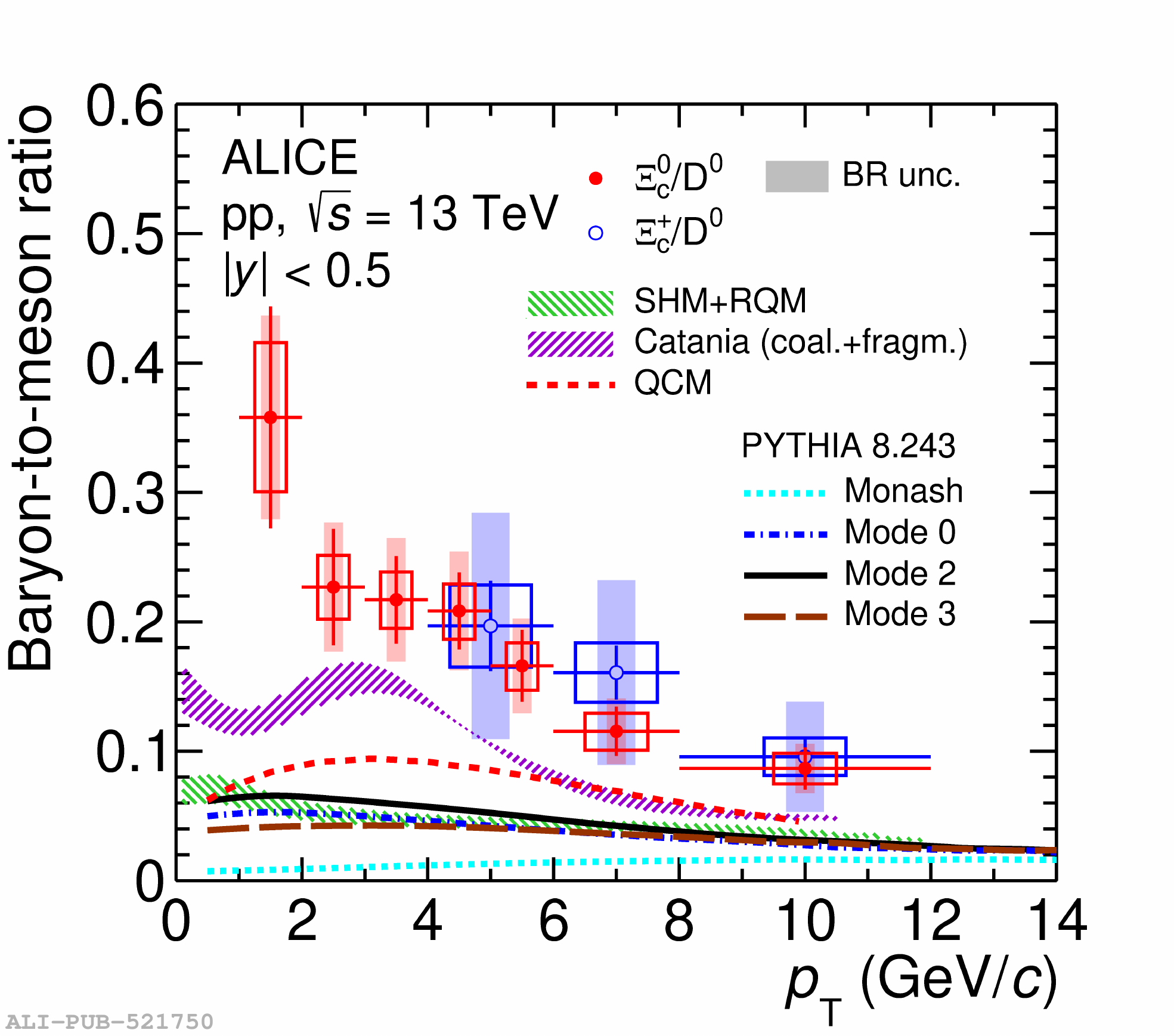}
\includegraphics[width=0.33\textwidth]{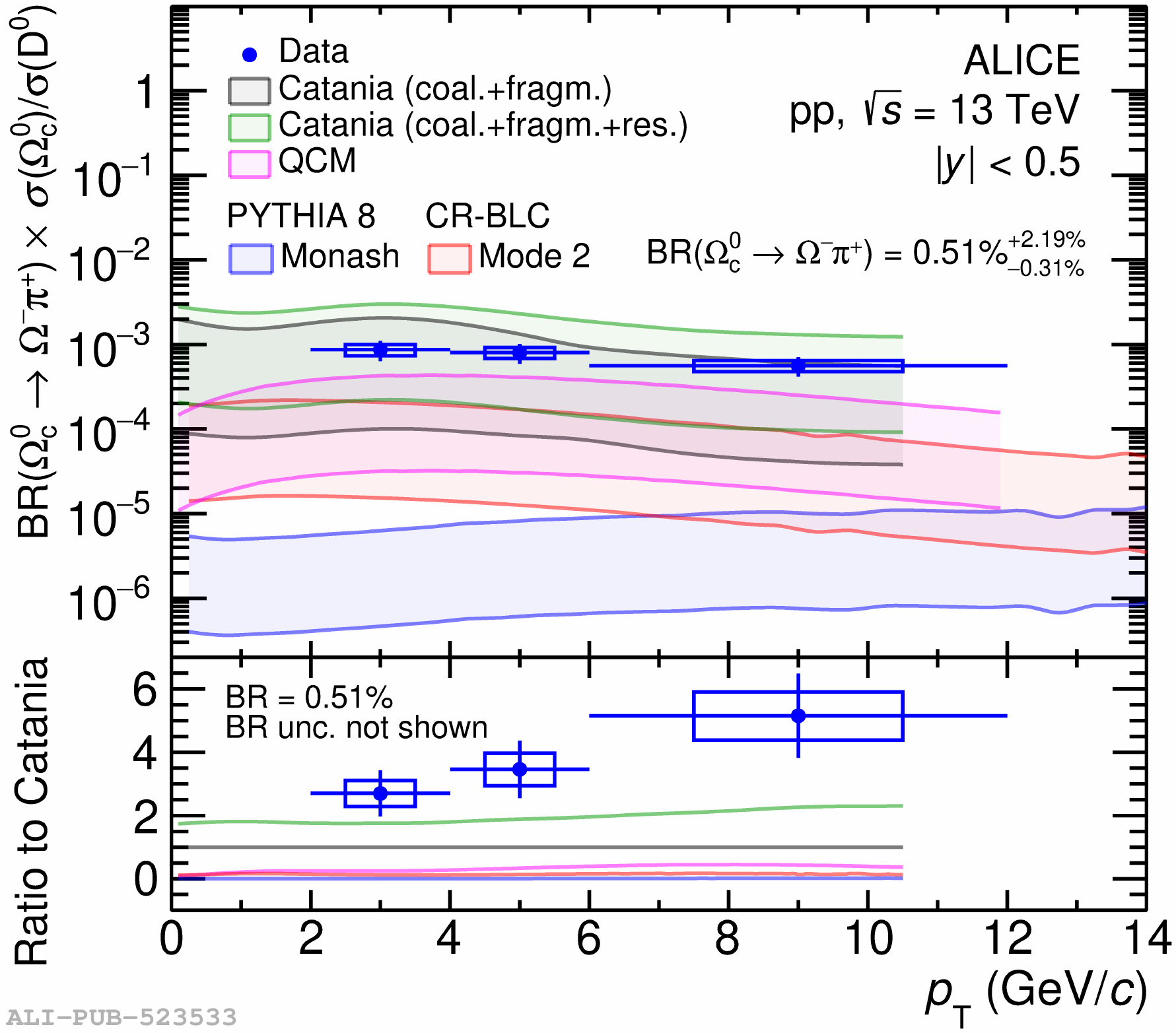}
\caption{Left: Prompt $\rm \Sigma_c^{0,+,++}/D^0$ yield ratio in pp collisions at $\sqrt{s}=13$ TeV \cite{ALICE:2020wla}. Middle: Prompt $\rm \Xi_c^0/D^0$ and $\rm \Xi_c^+/D^0$ yield ratios in pp collisions at $\sqrt{s}=13$ TeV \cite{ALICE:2021bli}. Right: Prompt $\rm BR\times\Omega_c^0/D^0$ yield ratio in pp collisions at $\sqrt{s}=13$ TeV \cite{ALICE:2022cop}. All the charm baryon-to-meson yield ratios are compared to theoretical calculations \cite{Skands:2014pea, Christiansen:2015yqa, He:2019tik, Song:2018tpv, Minissale:2020bif}.}
\label{fig_2}       
\end{figure}

The measurement of $\rm \Sigma_c^{0, +, ++}$ production is very important to constrain the production of $\rm \Lambda_c^+$ since $\rm \Sigma_c^{0, +, ++}$ almost completely decays to $\rm \Lambda_c^+$. In the left panel of Fig.~\ref{fig_2}, the $\rm \Sigma_c^{0, +, ++}/D^0$ yield ratio is shown as a function of $p_{\rm T}$ in pp collisions at $\sqrt{s}=13$~TeV and compared with model predictions. The $\rm \Sigma_c^{0, +, ++}/D^0$ yield ratio shows a similar $p_{\rm T}$ trend as the $\rm \Lambda_c^+/D^0$ yield ratio. The data indicate that the enhancement of $\rm \Lambda_c^+/D^0$ ratio in pp collisions can be partially explained by $\rm \Sigma_c^{0, +, ++}$ feed-down. All models mentioned above except PYTHIA 8 with Monash tune can describe the measured $\rm \Sigma_c^{0, +, ++}/D^0$ yield ratio.

The strange-charm $\rm \Xi_c^{0}$ and $\rm \Xi_c^{+}$ baryons are also investigated, and the $\rm \Xi_c^{0}/D^0$ and $\rm \Xi_c^{+}/D^0$ yield ratios measured in pp collisions at $\sqrt{s}=13$ TeV are shown in the middle panel of Fig.~\ref{fig_2}. In the right panel of Fig.~\ref{fig_2}, the measurement of the $\rm \Omega_c^0/D^0$ cross-section ratio times branching ratio (BR) of the $\rm \Omega_c^0\rightarrow\Omega^-\pi^+$ decay is reported for pp collisions at $\sqrt{s}=13$ TeV. The absolute BR of $\rm \Omega_c^0\rightarrow\Omega^-\pi^+$ is not measured hence, in order to compare data with models, a theoretical calculation of $\rm BR(\Omega_c^0\rightarrow\Omega^-\pi^+)=(0.51^{+2.19}_{-0.31})\%$, obtained by considering the estimate reported in Ref.~\cite{Hsiao:2020gtc} for the central value, and the envelope of the values (including their uncertainties) reported in Refs.~\cite{Hsiao:2020gtc, Gutsche:2018utw, Cheng:1996cs, Hu:2020nkg, Solovieva:2008fw, Wang:2022zja} to determine the uncertainty, is used to multiply different models. The $\rm \Xi_c^0/D^0$ and $\rm \Xi_c^+/D^0$ yield ratios show a decreasing trend with $p_{\rm T}$. The PYTHIA 8 Monash \cite{Skands:2014pea} tuned on measurements in $\rm e^+e^-$ collisions largely underestimates all charm baryon-to-meson yield ratios, providing evidence of different charm hadronization mechanisms between $\rm e^+e^-$ and pp collisions. The Catania model \cite{Minissale:2020bif} is the only one that provides a fair description of the $\rm \Xi_c^{0,+}/D^0$ and $\rm BR\times\Omega_c^0/D^0$ yield ratios. It is interesting to note that in this model the formation of a small-size quark--gluon plasma (QGP) in pp collisions is assumed and hadronization occurs via coalescence in addition to fragmentation.

\section{Prompt $\rm \Lambda_c^+/D^0$ yield ratio in p--Pb collisions}

\begin{figure}
\centering
\includegraphics[width=0.352\textwidth]{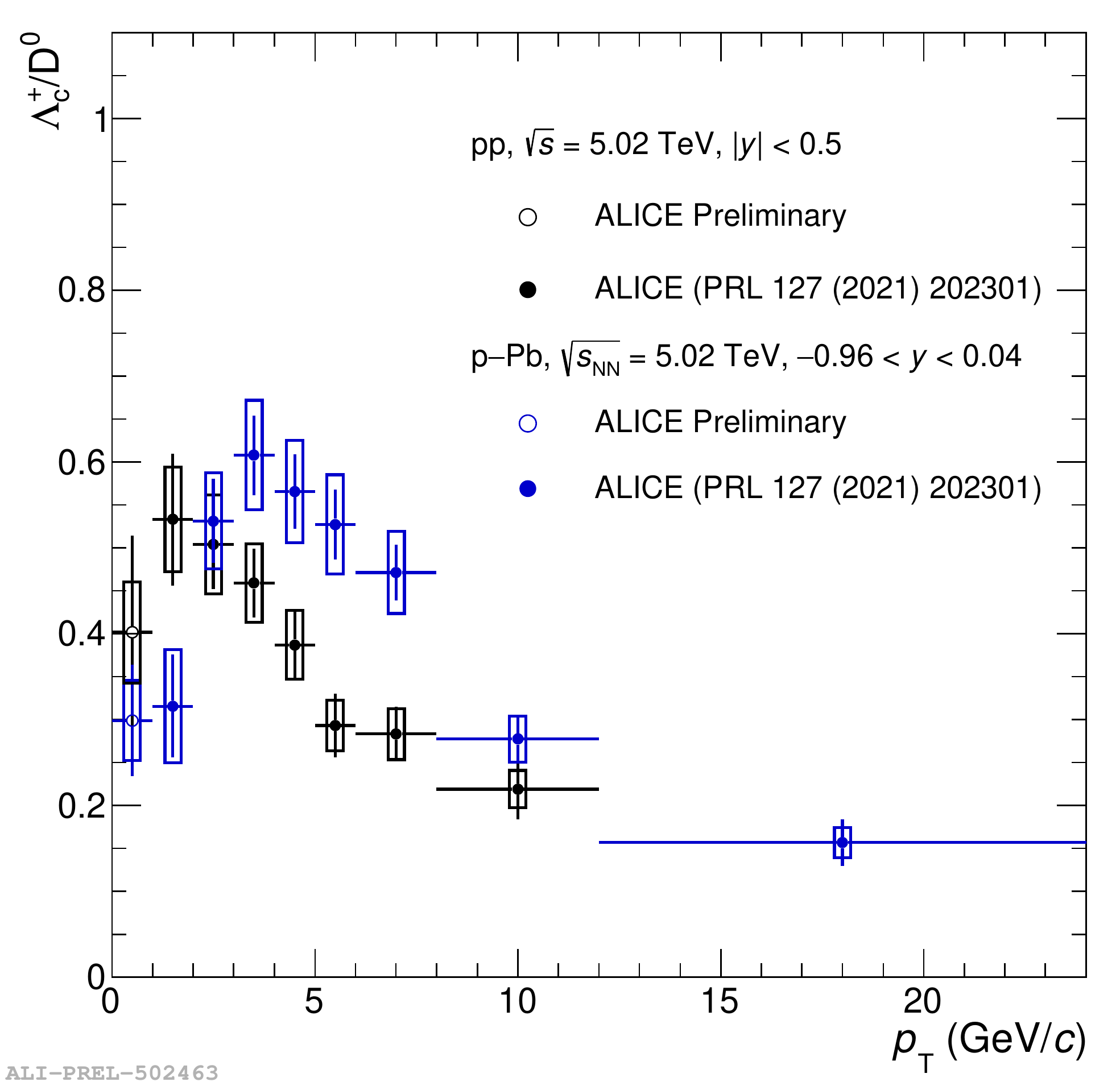}
\includegraphics[width=0.4\textwidth]{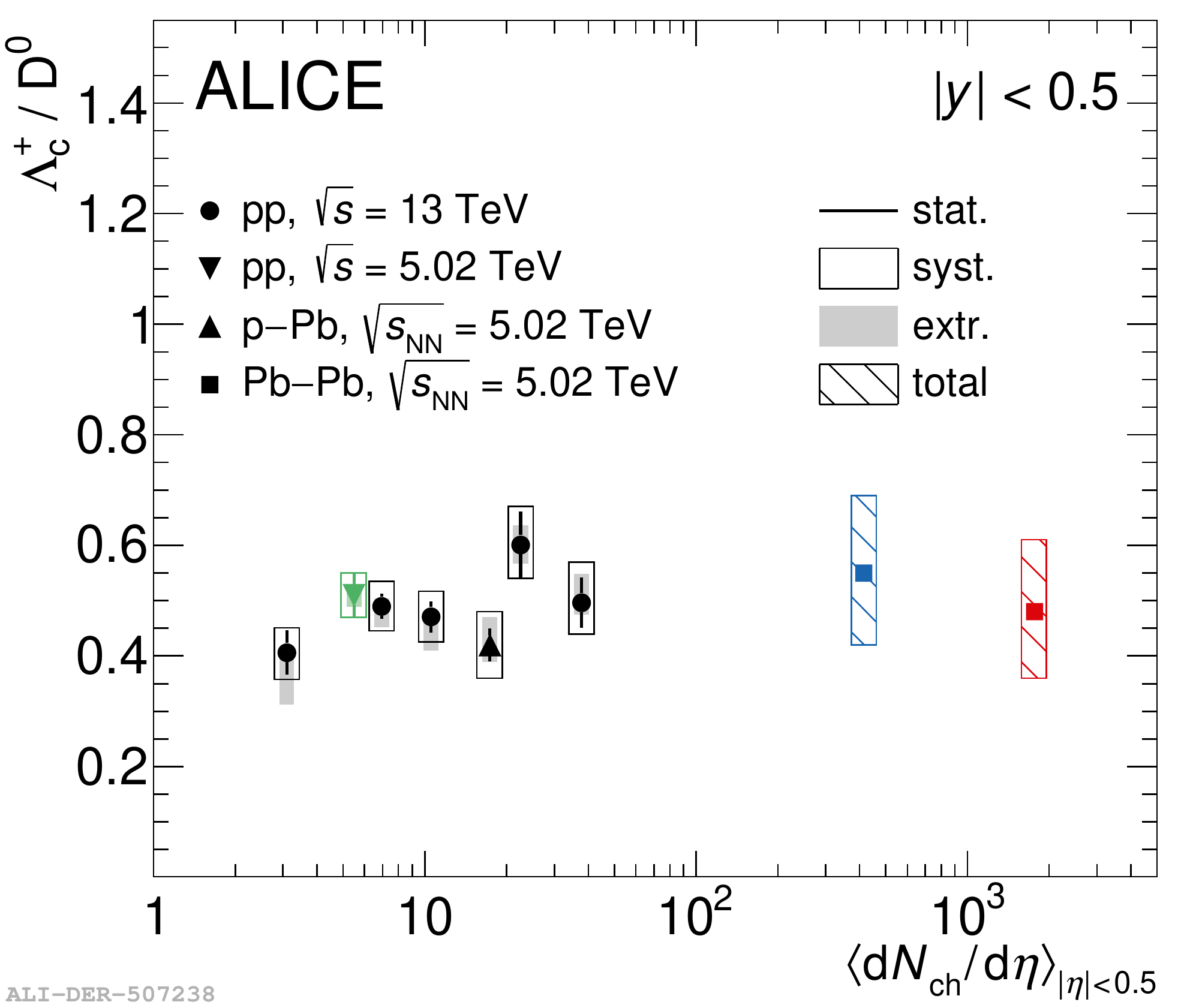}
\caption{Left: Prompt $\rm \Lambda_c^+/D^0$ yield ratio as function of $p_{\rm T}$ in pp and p--Pb collisions at $\sqrt{s_{\rm NN}}=5.02$~TeV. Right: $p_{\rm T}$-integrated prompt $\rm \Lambda_c^+/D^0$ yield ratio as a function of charged-particle multiplicity in pp, p--Pb and Pb--Pb collisions at various collisions energies \cite{ALICE:2021npz, ALICE:2021bib}.}
\label{fig_3}       
\end{figure}

The prompt $\rm \Lambda_c^+/D^0$ yield ratio measured down to $p_{\rm T}=0$ in p--Pb collisions at $\sqrt{s_{\rm NN}}=5.02$~TeV is shown in the left panel of Fig.~\ref{fig_3} and is compared with the measurement in pp collisions at $\sqrt{s}=5.02$~TeV. The p--Pb result for $p_{\rm T}>4$~GeV/$c$ suggests a larger ratio than that measured in pp collisions, while there is a hint of suppression in $p_{\rm T}<2$~GeV/$c$. This suggests a possible modification of $\rm D^0$ and $\rm \Lambda_c^+$ formation probabilities as a function of $p_{\rm T}$ in p--Pb collisions with respect to pp collisions, which could be attributed to a contribution of collective effects, i.e. radial flow, in p--Pb collisions.
The $p_{\rm T}$-integrated prompt $\rm \Lambda_c^+/D^0$ yield ratio as a function of charged-particle multiplicity is shown in the right panel of Fig.~\ref{fig_3}. There is no significant variation as a function of multiplicity, collision system, or collision energy, which indicates that the modification observed in the study of $p_{\rm T}$ differential yields could be due to a momentum redistribution without a modification of the overall $\rm \Lambda_c^+$ yield relative to the $\rm D^0$ one.

\section{Charm production and fragmentation in pp and p--Pb collisions}
\begin{figure}
\centering
\includegraphics[width=0.4\textwidth]{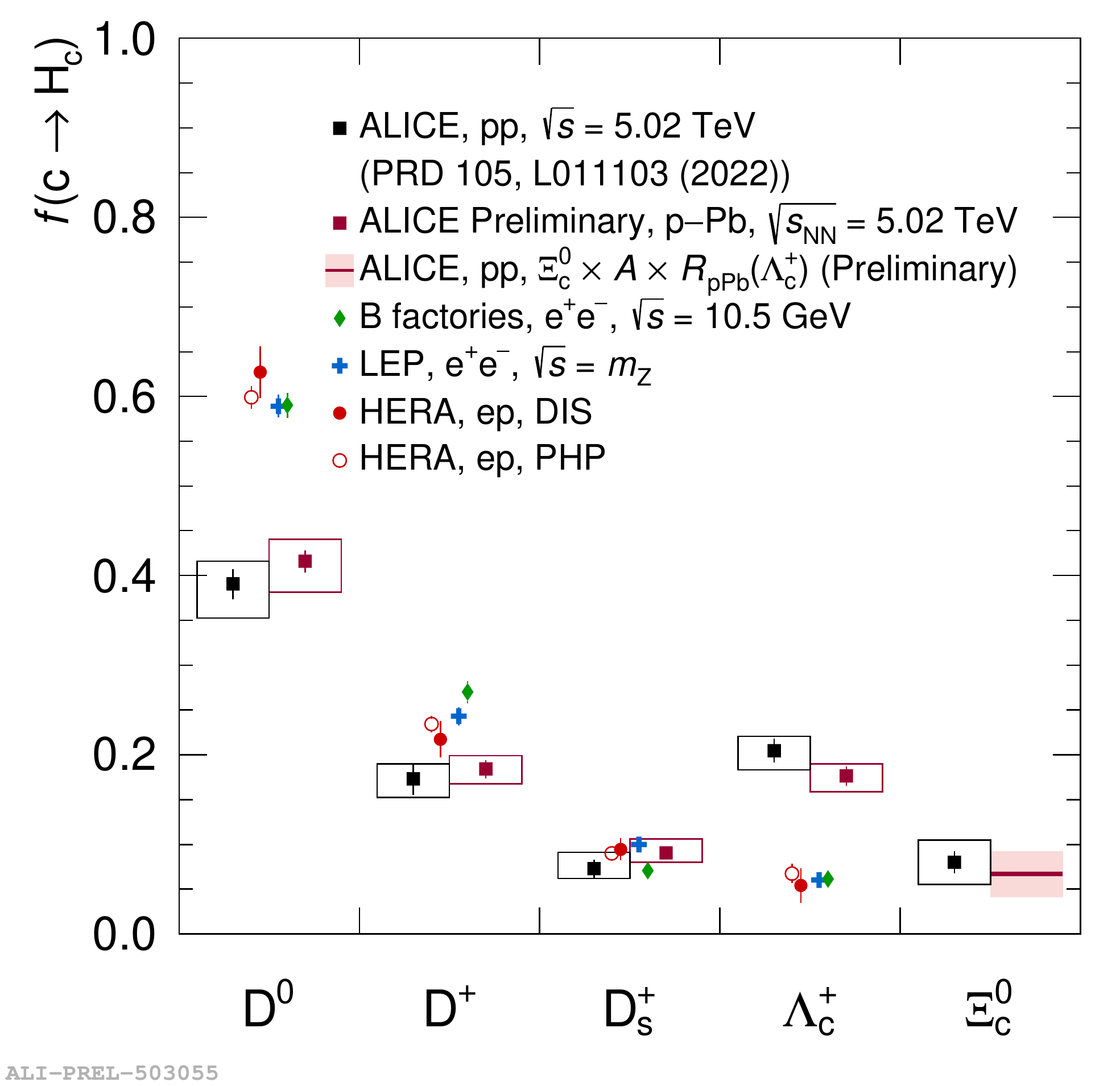}
\includegraphics[width=0.4\textwidth]{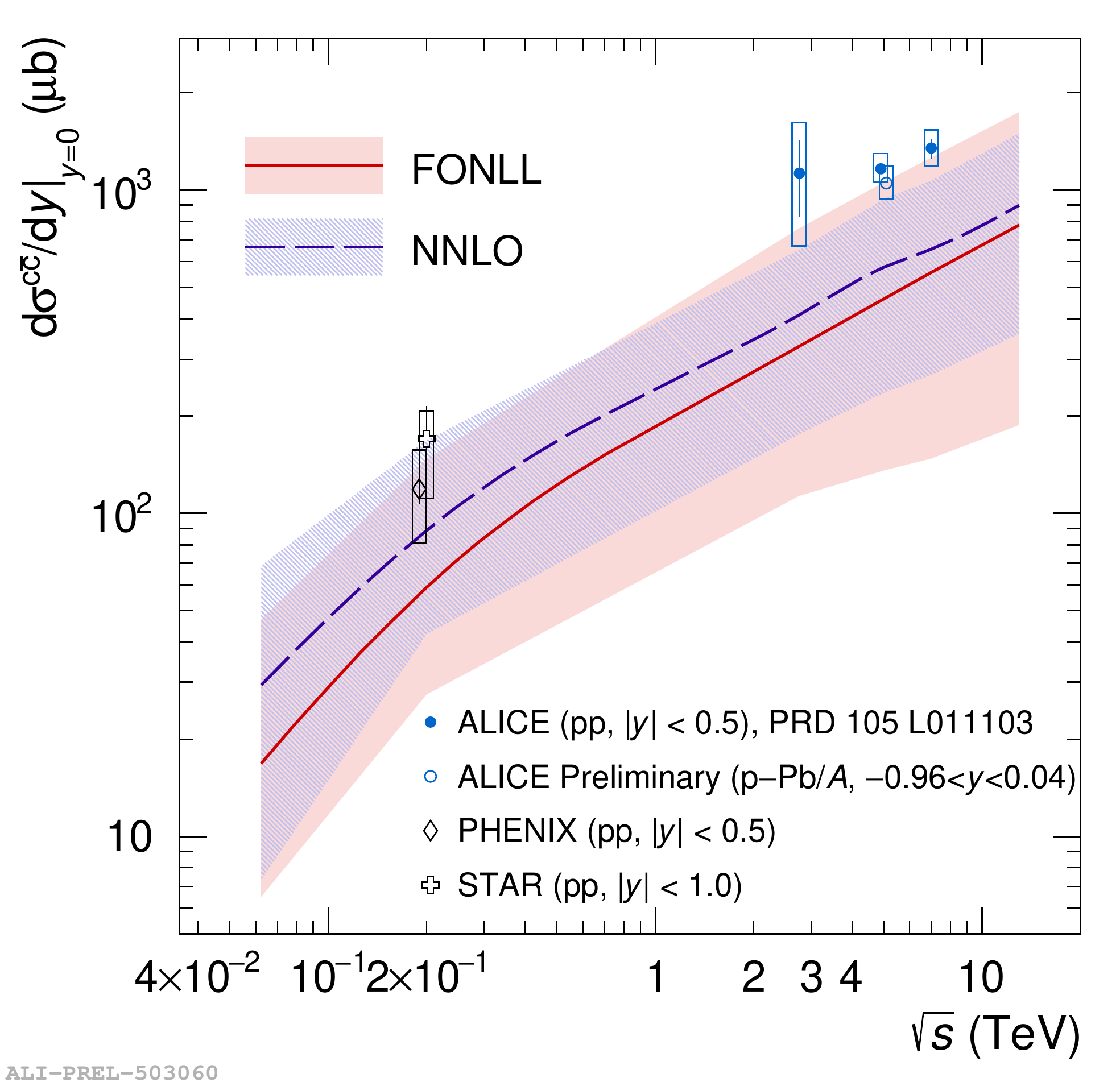}
\caption{Left: Charm-quark fragmentation fractions into charm hadrons measured in pp and p--Pb collisions at $\sqrt{s_{\rm NN}}=5.02$~TeV in comparison with experimental measurements performed in $\rm e^+e^-$ and ep collisions \cite{ALICE:2021dhb}. Right: Charm production cross section at midrapidity per unit of rapidity as a function of the collision energy at the LHC \cite{ALICE:2021dhb} and RHIC \cite{STAR:2012nbd, PHENIX:2010xji} compared to FONLL \cite{Cacciari:2012ny} and NNLO \cite{dEnterria:2016ids} calculations.}
\label{fig_4}       
\end{figure}

The charm fragmentation fraction $f(\rm c\rightarrow H_c)$ shown in the left panel of Fig.~\ref{fig_4} represents the probability of a charm quark hadronizing into a given charm hadron species. The fraction of charm quarks that hadronize into baryons is about 40\%, which is four times larger than what was measured at colliders with electron beams, showing that the assumption of the charm fragmentation universality (collision-system independence) is not valid.

The $\rm c\bar{c}$ production cross section per unit of rapidity at midrapidity ($\rm d{\it \sigma}^{c\bar{c}}/d{\it y}|_{|{\it y}| < 0.5}$) is calculated by summing the $p_{\rm T}$-integrated cross sections of all measured ground-state charm hadrons ($\rm D^0$, $\rm D^+$, $\rm D_s^+$, $\rm \Lambda_c^+$, $\rm \Xi_c^0$ and their charge conjugates). The contribution of $\rm \Xi_c^0$ is multiplied by a factor of 2 in order to account for the contribution of $\rm \Xi_c^+$. Due to the absence of a $\rm \Omega_c^0$ production measurement in pp collisions at $\sqrt{s}=5.02$ TeV, an asymmetric systematic uncertainty is assigned assuming a contribution equal to the one of $\rm \Xi_c^0$ considering the prediction of the Catania model \cite{Minissale:2020bif}. The resulting $\rm c\bar{c}$ production cross section per unit of rapidity at midrapidity in p--Pb collisions after scaling for Pb ion mass number, shown in the right panel of Fig.~\ref{fig_4} together with measurements at RHIC \cite{STAR:2012nbd, PHENIX:2010xji}, is compatible with that in pp collisions. The fragmentation fractions obtained in pp collisions at $\sqrt{s}=5.02$ TeV allow the recomputation of the charm production cross sections per unit of rapidity at midrapidity in pp collisions at $\sqrt{s}=2.76$ and 7 TeV, which are about 40\% higher than the previously published results \cite{ALICE:2012inj, ALICE:2017olh} which used the D-meson fragmentation fraction from $\rm e^+e^-$ collisions. The measured $\rm c\bar{c}$ production cross sections per unit of rapidity at midrapidity are located at the upper edge of FONLL \cite{Cacciari:2012ny} and NNLO \cite{dEnterria:2016ids} predictions.

\section{Acknowledgments}
This work was supported by the National Natural Science Foundation of China (NSFC) (No. 12105109).

\let\oldthebibliography\thebibliography
\let\endoldthebibliography\endthebibliography
\renewenvironment{thebibliography}[1]{
  \begin{oldthebibliography}{#1}
    \setlength{\itemsep}{0.2em}
    \setlength{\parskip}{0em}
}
{
  \end{oldthebibliography}
}

\bibliographystyle{utphys}
\bibliography{references}

\end{document}